\documentclass[twocolumn, times, trackchanges]{aastex63}

\usepackage{amsmath}
\usepackage{txfonts}
\usepackage{graphicx}
\graphicspath{{./}{figures/}}
\newcommand{\target}{NGC\,2355 }

\begin{document}

\title{Searching for Variable Stars in the Open Cluster \target and Its Surrounding Region}
\shorttitle{Variable Stars in \target}

\shortauthors{Wang, H. et al.}

\author[0000-0001-8026-787X]{Hong Wang}
\affiliation{Xinjiang Astronomical Observatory, Chinese Academy of Sciences, Urumqi, Xinjiang 830011, People's Republic of China}
\affiliation{School of Astronomy and Space Science, University of Chinese Academy of Sciences, Beijing 100049, People's Republic of China}

\author[0000-0001-7134-2874]{Yu Zhang}
\email{zhy@xao.ac.cn}
\affiliation{Xinjiang Astronomical Observatory, Chinese Academy of Sciences, Urumqi, Xinjiang 830011, People's Republic of China}
\affiliation{School of Astronomy and Space Science, University of Chinese Academy of Sciences, Beijing 100049, People's Republic of China}

\author[0000-0002-8049-202X]{Xiangyun Zeng}
\affiliation{Center for Astronomy and Space Sciences, China Three Gorges University, Yichang 443000, People's Republic of China}

\author[0000-0002-5448-6267]{Qingshun Hu}
\affiliation{Xinjiang Astronomical Observatory, Chinese Academy of Sciences, Urumqi, Xinjiang 830011, People's Republic of China}
\affiliation{School of Astronomy and Space Science, University of Chinese Academy of Sciences, Beijing 100049, People's Republic of China}

\author[0000-0002-7420-6744]{Jinzhong Liu}
\affiliation{Xinjiang Astronomical Observatory, Chinese Academy of Sciences, Urumqi, Xinjiang 830011, People's Republic of China}
\affiliation{School of Astronomy and Space Science, University of Chinese Academy of Sciences, Beijing 100049, People's Republic of China}

\author[0000-0002-0040-8351]{Mingfeng Qin}
\affiliation{Xinjiang Astronomical Observatory, Chinese Academy of Sciences, Urumqi, Xinjiang 830011, People's Republic of China}
\affiliation{School of Astronomy and Space Science, University of Chinese Academy of Sciences, Beijing 100049, People's Republic of China}

\author[0000-0002-3839-4864]{Guoliang L$\rm{\ddot{u}}$}
\affiliation{School of Physical Science and Technology, Xinjiang University, Urumqi, 830064, People’s Republic of China}

\begin{abstract}
We have investigated the variable stars in the field surrounding \target based on the time-series photometric observation data. More than $3000$ CCD frames were obtained in the $V$ band spread over $13$ nights with the Nanshan One-meter Wide-field Telescope. We have detected $88$ variable stars, containing $72$ new variable stars and $16$ known variable stars. By analyzing these light curves, we classified the variable stars as follows: $26$ eclipsing binaries, $52$ pulsating stars, $4$ rotating variables, and $6$ unclear type variable stars for which their periods are much longer than the time baseline chosen. Employing {\it Gaia} DR2 parallax, kinematics, and photometry, the cluster membership of these variable stars were also analyzed for NGC\,2355. In addition to the $11$ variable members reported by \citet{18cant}, we identify $4$ more variable member candidates located at the outer region of \target and showed homogeneity in space positions and kinematic properties with the cluster members. The main physical parameters of \target estimated from the two-color and color-magnitude diagrams are log(age/yr) = 8.9, $E(B - V) = 0.24$ mag, and [Fe/H] = - 0.07 dex. 
\end{abstract}


\section{Introduction}\label{sec:Int}
Open clusters are exceptional laboratories for us to understand the fundamental astrophysical processes and one of the principal components of the Galaxy. It is believed that a significant number of stars in the Galaxy are formed in clusters \citep{20josh}. Therefore, their extensive study is vital to infer the star formation history in the Galaxy \citep{94phel} and to probe the Galactic structure \citep{03chen, 18cant}. Chemical abundance, age, spatial distribution, and kinematic characteristics of open clusters are the main keys to studying the formation and evolution of stars, clusters, and Galactic disc. Generally, most of the stars in the clusters exist in a variational form, and their luminosity changes chiefly for their internal reasons as well as external factors. The study of variable stars can provide a great deal of extra information that allows us to tighten the screws on theoretical models attempting to predict stellar properties, such as mass, radius, brightness, temperature, internal and external structure, some of which are difficult or impossible to obtain in stars other than variable stars \citep{06pisk, 11sand}.

The open cluster \target [RA(J2000.0) = $07^{h}16^{m}59.3^{s}$, Dec(J2000.0) = $13^{\circ}46^{'}19^{''}$] is situated in the anti-center direction of the Galaxy, in the constellation of Gemini. In one of the earliest studies of NGC\,2355, a photometric survey in $U, B, V$ bands down to $V$ $\sim$ 18.1 mag was taken by \citet{91kalu} with the 0.9-m telescope at Kitt Peak National Observatory and they estimated the reddening of \target to be $E(B-V)$ = 0.12 mag, the distance modulus $(m-M)_0$ = 12.1, and the metallicity was +0.13 dex. \citet{99ann} examined open cluster \target as a part of the Bohyunsan Optical Astronomy Observatory photometric survey in Korea and determined the [Fe/H] = - 0.32 dex, $E(B - V)$ = 0.25 mag, $(m - M)_0$ = 11.4 mag and age of 1 Gyr from $U, B, V, I$ photometry by Doyak 1.8m telescope subsequently. Based on the two previously mentioned studies, \citet{13oliv} used a method for estimating the metallicity of open clusters via non-subjective isochrone fitting using the cross-entropy global optimization algorithm applied to $U, B, V$ photometric data. There was a disagreement in their work due to the two different extinction groups, the former with the age of 0.8 Gyr, the reddening of 0.22 mag, and the metallicity of -0.23 dex, while the latter with the age of 0.9 Gyr, $E(B - V)$ = 0.32 mag and [Fe/H] = - 0.32 dex. More recently, high-resolution spectra had been obtained for particular stars in or around the open cluster \target by \citet{00soub}, \citet{11jaco}, and \citet{15dona}, and more precise metallicity were derived, approximately - 0.06 to - 0.08 dex. 

However, to our knowledge, few studies researched the detailed variables for the open cluster NGC\,2355, but only at the stage of photometry and spectroscopy for the cluster properties and parameters. This also provides an additional opportunity to characterize variable stars in this open cluster. In our work, we carry out an extensive search for variable stars in NGC\,2355. The framework of this paper is structured as follows: observations and the data reduction procedures are presented in Section~\ref{sec:O&D}. We describe the circumstantial classification of variable stars in Section~\ref{sec:varible}. In Section~\ref{sec:disc}, we focus on the membership of the variable member stars and the stellar parameters associated with this open cluster, and we have also discussed the possibility of variable stars being missing cluster members. We present a summary of results in Section~\ref{sec:SM}.

\section{observations and data reductions}\label{sec:O&D}
All of the photometric observations for our study were taken with the Nanshan One-meter Wide-field Telescope (NOWT) at the Nanshan station of the Xinjiang Astronomical Observatory. The camera is equipped with an E2V $4160 \times 4136$ CCD203-82 with the pixel size of 12 $\mu$m near the center of the CCD clip, yielding a scale of 1.125$^{''}$ per pixel and corresponding to a $78 \times 78$ arcmin$^{2}$ field of view around the center of the open cluster NGC\,2355. The CCD operates at about - 120 $^{\circ}$C with liquid nitrogen cooling thus the dark current is less than 1 $e^{-}$ pix$^{-1}$ h$^{-1}$ at - 120 $^{\circ}$C. To identify variable stars in NGC\,2355, time-series observations were taken in the $V$ band of the Johnson-Cousin-Bessel during $13$ nights from January 8 to 10 in 2018 and December 10 to 19 in 2020 \citep{53john,76cous,90bess}. The detailed log of the observations is listed in Table~\ref{tab:log}. 
To secure photometric calibration of the CCD system, the standard star fields $J005945 + 440830$ and $J234139 + 453900$ \citep{16clem}, together with \target were also observed in $U, B, V, R, I$ bands on the night of 2018 January 9, as summarized in Table~\ref{tab:stand}. Several biases and twilight flat-field frames were also taken in $U, B, V, R$, and $I$ during the observing night.

\begin{deluxetable*}{cccccccc}[!htp]
\tabletypesize{\small}
\tablewidth{0pc}
\tablenum{1}
\tablecaption{Log of observations of \target and its adjacent sky region run in $V$ band.\label{tab:log}}
\tablehead{
\colhead{Date} & \colhead{CCD} & \colhead{FOV} & \colhead{Length} & \colhead{Frames} & \colhead{Exposure Time} & \\
\colhead{(yyyy/mm/dd)} & \colhead{(pixel$^{2}$)} & \colhead{(arcmin$^{2}$)} & \colhead{(h)} & \colhead{$(V~band)$} & \colhead{(s)} & 
}
\startdata
2018 Jan 08 & $4160 \times 4136$ & $78 \times 78$ & 8 & 369 &35 \\
2018 Jan 09 & $4160 \times 4136$ & $78 \times 78$ & 8 & 360 &35 \\
2018 Jan 10 & $4160 \times 4136$ & $78 \times 78$ & 8 & 308 &35 \\
2020 Dec 10 & $4160 \times 4136$ & $78 \times 78$ & 7 & 202 &30 \\
2020 Dec 11 & $4160 \times 4136$ & $78 \times 78$ & 8 & 358 &30 \\
2020 Dec 12 & $4160 \times 4136$ & $78 \times 78$ & 8 & 341 &30 \\
2020 Dec 13 & $4160 \times 4136$ & $78 \times 78$ & 8 & 357 &30 \\
2020 Dec 14 & $4160 \times 4136$ & $78 \times 78$ & 8 & 354 &30 \\
2020 Dec 15 & $4160 \times 4136$ & $78 \times 78$ & 5 & 105 &30 \\
2020 Dec 16 & $4160 \times 4136$ & $78 \times 78$ & 3 & 81 &30 \\
2020 Dec 17 & $4160 \times 4136$ & $78 \times 78$ & 3 & 92 &30 \\
2020 Dec 18 & $4160 \times 4136$ & $78 \times 78$ & 2 & 69 &30 \\
2020 Dec 19 & $4160 \times 4136$ & $78 \times 78$ & 2 & 68 &30 \\
\enddata
\end{deluxetable*}

\begin{deluxetable}{ccccccc}[!htp]
\tabletypesize{\small}
\tablewidth{0pc}
\tablenum{2}
\tablecaption{$UBVRI$ multi-color photometric observation of \target and two standard star fields in 2018 January 9.\label{tab:stand}}
\tablehead{
\colhead{Target} & \colhead{$U$} & \colhead{$B$} & \colhead{$V$} & \colhead{$R$} & \colhead{$I$} &
}
\startdata
J005945+440830 & $240(s)\times3$  & $90(s)\times3$ & 60(s)$\times3$ & $50(s)\times3$ & $80(s)\times3$ \\
J234139+453900 & $240(s)\times3$  & $90(s)\times3$ & 60(s)$\times3$ & $50(s)\times3$ & $80(s)\times3$ \\
\target & $240(s)\times3$  & $90(s)\times3$ & 60(s)$\times3$ & $50(s)\times3$ & $80(s)\times3$ \\
\enddata
\end{deluxetable}

The IRAF \footnote{Image Reduction and Analysis Facility, is distributed by the National Optical Astronomy Observatory, which is operated by the Association of Universities for Research in Astronomy, Inc., under cooperative agreement with the National Science Foundation. \url{https://iraf-community.github.io/}} package was used for pre-processing of data frames which include bias subtraction and flat-field correction \citep{86tody,93tody}. To identify objects in the CCD frames, the pixel coordinates of the frames were converted into equatorial coordinates by matching with the third US Naval Observatory CCD Astrograph Catalog (UCAC3). For all of the CCD frames we obtained, the photometry was carried out by SExtractor \footnote {SExtractor, \url{https://www.astromatic.net/software/sextractor/} \citep{96bert}.} SExtractor is a program that can perform reasonably well on astronomical images of moderately crowded star fields such as open clusters. It uses the K-$\sigma$ Clipping method to calculate the distribution of background values with the location of all small areas in CCD frames. For crowded star fields, it performs median filtering on the distribution to suppress possible local overestimations due to bright stars \citep{96bert}. Based on the IRAF and SExtractor, a pipeline is built, for data processing, to obtain the instrumental magnitudes that are used widely for NOWT time-domain surveys \citep{16song, 18ma, 21li}.

\subsection{Photometric calibration} \label{sub:calib}
The standard fields $J005945 + 440830$ and $J234139 + 453900$ \citep{16clem} were observed during one observing night in order to calculate the instrumental photometry to the standard $U, B, V, R ,I$ photometric system.  Our $UBVRI$ data were calibrated through about 69 standard stars in \citet{16clem} field $J005945 + 440830$. 
It is generally believed that the atmospheric extinction coefficient of the same site remains basically stable within a certain time interval, so for them, we assumed the typical values for the Nanshan site \citep{20bai}. The resulting transformation equations are:

\begin{equation}
\begin{aligned}
u=&U+(4.820\pm0.023)+(0.590\pm0.022)\times X+\\
&(-0.156\pm0.061)\times (U-B)\\
b=&B+(1.810\pm0.017)+(0.431\pm0.029)\times X+\\
&(-0.066\pm0.021)\times (B-V)\\
v=&V+(2.107\pm0.012)+(0.282\pm0.026)\times X+\\
&(0.125\pm0.016)\times (B-V)\\
r=&R+(2.125\pm0.015)+(0.217\pm0.019)\times X+\\
&(0.238\pm0.031)\times (V-I)\\
i=&I+(3.210\pm0.011)+(0.156\pm0.021)\times X+\\
&(-0.022\pm0.014)\times (V-I),
\end{aligned}
\label{eq:UBVRI}
\end{equation}
where $u, b, v, r, i$ represent the instrumental magnitudes, $U, B, V, R, I$ stand for the magnitudes in the standard system, X is for the airmass, and the remaining two terms are the constant and the color.  

\begin{figure}[!htp]
\begin{center}
\includegraphics[angle=0, width=0.7\columnwidth]{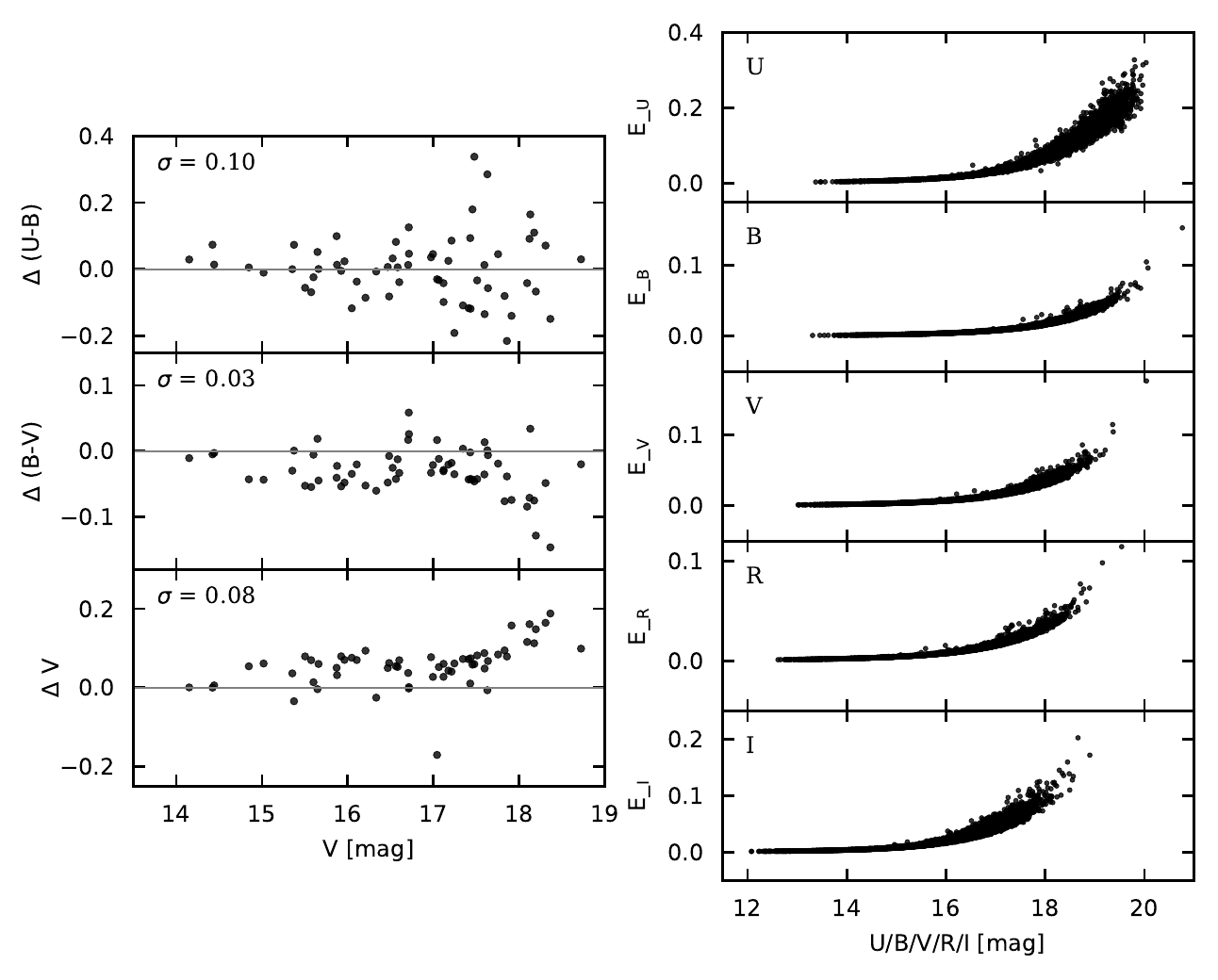}
\caption{Left panels: Comparisons of the photometry presented in this study with that of \citet{16clem} for field $J234139 + 453900$ in $V$ magnitudes and ($U-B$) -- ($B-V$) colors. Right panels: Photometric errors against its corresponding $U,B,V,R,$ and $I$ magnitudes on the standard system from top to bottom. }
\label{fig:err}
\end{center}
\end{figure}

According to the transformation Equation \ref{eq:UBVRI} and the observation of the standard field $J234139 + 453900$, we calculate the $UBVRI$ standard magnitudes for stars in the standard field $J234139 + 453900$ to assess the overall quality of the calibrated magnitudes and colors. By cross-correlating with the standard magnitudes of \citet{16clem}, a total of 65 standard stars are found in common. The comparisons of $V$ magnitudes, ($U-B$) and ($B-V$) colors between these two catalogues are shown in the left panels of Figure \ref{fig:err}. As illustrated, the final standard deviations are $0.08, 0.03, 0.1$ in $V$, ($B-V$), and ($U-B$), respectively. This indicates that our $V$, ($B-V$), and ($U-B$) measurements are in fair agreement with those given in \citet{16clem}. We present the photometric error derived from photometric reduction plotted against its corresponding $U,B,V,R,$ and $I$ magnitudes in the right panels of Figure \ref{fig:err}.

\subsection{Variable star Detection} \label{sub:detec}
For the light curves of the stars observed in the field of view, we use the SysRem algorithm for systematic error subtraction \citep{05tamu,10ofir}. The clouds passing through the field of view will affect the flux measurement of some stars and the observation effect changes over time, so least-squares fitting is employed to perform a coarse initial decorrelation. Least-squares fitting for the initial zero-point correction is necessary for a better variables searching during the time-domain observations. The detrending algorithm SysRem is employed to eliminate systematic effects hidden in a large set of photometric light curves. Nevertheless, there are additional variances remaining in the corrected data set. Then we use the Gaussian Mixed Model estimation method to determine additional variances in the data set after correction \citep{20gaox,22luox}. After several repeated iterations of the algorithm, we can get the different light curves of all the stars in the field of view. By visual inspection, we can obtain a number of variable star candidates easily, which can then be analyzed in detail later.

The differential magnitude of each star is given by \citet{06coll} as: 
\begin{equation}
 x_{ij}=m_{ij}-\hat{m}_{j}-\hat{z}_{i}, 
\label{eq:LebsequeI}
\end{equation}
where $m_{ij}$ is a two-dimensional array of instrumental magnitude, the index $i$ denotes a single CCD frame while the entire seasons' data, the second index $j$ labels an individual star. $\hat{m}_{j}$ is the mean instrument magnitude for each numbered star and $\hat{z}_{i}$ is the zero-point correction for each CCD frame. After repeated iterations, the differential light curves of the stars in the field of view are obtained.

The software PERIOD04 \citep{05lenz} is employed to analyze the period of variable stars. The rectified light curve was fitted with the following formula:
\begin{equation}
m=m_{0} + \sum_{i=1}^{n} A_{i} \sin (2\pi (f_{i}t+\phi_{i})), \label{period}
\end{equation}
where $m_{0}$, $A_{i}$, $f_{i}$, and $\phi_{i}$ are zero-point, amplitude, frequency, and the corresponding phase, respectively.

This software adopts single-frequency Fourier and multi-frequency nonlinear least-squares fitting algorithms. It offers tools to extract the individual frequencies from the multi-periodic content of time series and provides a flexible interface to perform multiple-frequency fits \citep{05lenz}. The frequencies of the intrinsic and statistically significant peaks in the Fourier spectra can be extracted via iterative pre-whitening. The frequency investigations were stopped when the signal-to-noise (S/N) value is less than 4.0 \citep{93breg}. Amplitude spectra are computed after pre-whitening. The amplitude spectra of these stars are regarded as another criterion for finding variable stars. In our study, to determine the type of each variable star exactly, we use the main frequency with the highest signal-to-noise ratio extracted from the PERIOD04 to obtain the period of the pulsating variable star as a reference.

\subsection{{\it Gaia} and LAMOST Data} \label{sub:gala}
To identify the variable member stars for \target, we use the member catalogue of open clusters \citep{18cant}, which provides the combined spatio-kinematic-photometric membership of 1229 open clusters based on the membership assignment code Unsupervised Photometric Membership Assignment in Stellar Clusters (UPMASK) method \citep{14kron} applied on the {\it Gaia} DR2 \citep{18evan}. They identified 328 members of the open cluster NGC\,2355. Considering the precision of the data, the sources they used are $G < 18$ mag. The membership probabilities of these stars were obtained after 10 iterations of UPMASK, and the membership probability of a star corresponds to the frequency with which it was considered as a member by UPMASK \citep{14kron}. To get the cluster memberships for detected variable stars, we match the equatorial coordinates of more than 30,000 stars we observed with that of the 328 reported members within $1^{''}$. As a result, all of the cluster members with $V$ band photometric data are detected in our field of view. 

Large Sky Area Multi-Object Fiber Spectroscopic Telescope (LAMOST) low-resolution spectra have a resolution $R \sim 1800$ and cover the wavelength range $370-900$ nm \citep{12cui}. The spectral data released by LAMOST are selected based on the signal-to-noise ratio (S/N), generally those with S/N $\geqslant 15$ for bright night and S/N $\geqslant 6$ for dark night are selected, and those that do not meet the requirements are eliminated. These spectral data that satisfy the requirements are used to obtain the stellar atmospheric parameters for AFGK-type stars \citep{15luo}. The seventh data release of LAMOST Galactic Survey (LMOST DR7) has released in 2021, which contains more than 10 million stellar spectra \citep{12zhao}. With the importance of stellar parameters (e.g. $T_{eff}$) of determining variable star types, we cross-match with the low-resolution spectra catalogues of LAMOST DR7 to add extra information to the variable stars, aiming to make our variable star classification more comprehensive.  

\section{Variable Stars}\label{sec:varible}
\begin{figure}[!htp]
\begin{center}
\includegraphics[angle=0, width=0.8\columnwidth]{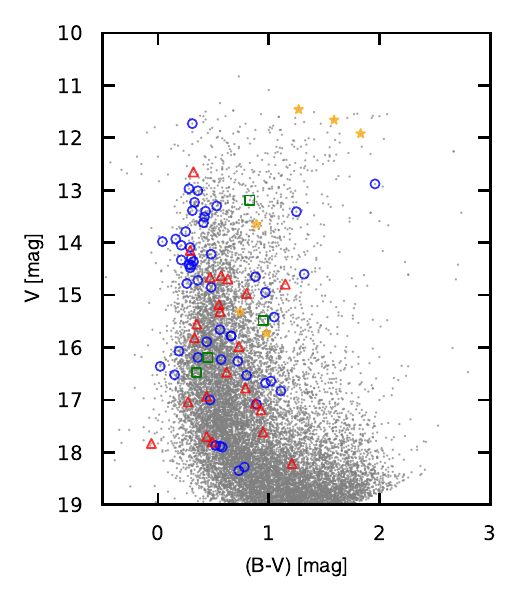}
\caption{Color-magnitude diagram for the observation filed with variables. The red, blue, green and orange symbols mark the eclipsing binaries, pulsators, rotating variables and unclear variable stars, respectively.}
\label{fig:VBV}
\end{center}
\end{figure}

Relating to the methods mentioned in Section~\ref{sec:O&D}, we first use visual inspection to identify variable stars such as eclipsing binaries and long-term variable stars, which can be directly distinguished. Others can be considered as variable stars when their light curves show significant periodic inherent changes in $V$ bands. In our work, we detect $88$ variable stars totally, we cross-match them with LAMOST DR7 and we can know that $22$ of them have spectral data, and we extract their temperatures as a reference. The information of periods, amplitudes, the shapes of phased light curves, and effective temperatures are taken to classify their types. As a result, we have identified $26$ eclipsing binaries, $52$ pulsators, $4$ rotating variable stars, and $6$ unclear type variables during the $88$ variables. And Figure \ref{fig:VBV} present the color-magnitude diagram (CMD) for these variable stars and the stars in the observed field. We then compare our classifications with other surveys, i.e., The All-Sky Automated Survey for Supernovae (ASAS-SN\footnote{\url{https://www.aavso.org/vsx/}}) and The International Variable Star Index (VSX\footnote{\url{https://asas-sn.osu.edu/variables/}}). As stated by \citet{18jaya} in the ASAS-SN survey description, the VSX catalogue seems to be the most comprehensive star catalogue of investigated variable stars in modern era, and includes many kinds of variable stars that detected by a great deal of optical surveys. Therefore, we matched $16$ known variable stars that had been determined and studied mainly by \citet{18jaya}, \citet{18chen} and \citet{20chen}. 

\subsection{Eclipsing Binaries} \label{sub:EB}
\begin{figure*}[!htp]
\begin{center}
\includegraphics[angle=0, width=1.8\columnwidth]{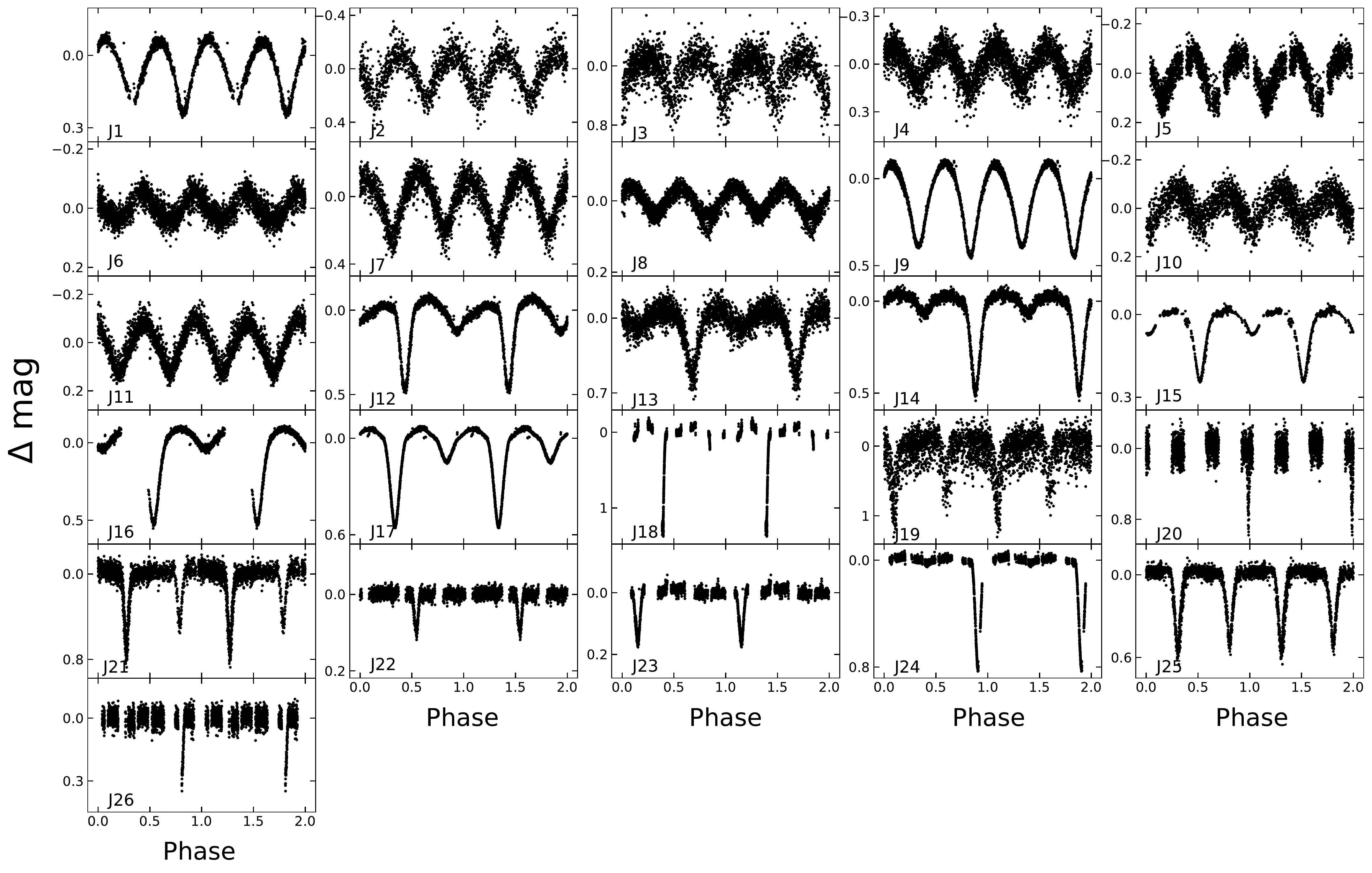}
\caption{The phase-folded light curves for these 26 eclipsing binaries.}
\label{fig:EB}	
\end{center}
\end{figure*}

We have detected 26 eclipsing binaries, the details are presented in Table~\ref{tab:EB}. 
For the amplitudes of these eclipsing binaries, we use PERIOD04 to obtain the periods of them for phase folding firstly, and then sort the phase data in the order from 0 to 1. Secondly, the mathematical method of moving average \citep{22shan} is employed to average the sorted data by taking 20 data in order from the first data. Finally, the amplitude of each eclipsing binary is obtained by subtracting the minimum value from the maximum value obtained. There are 11 EW-type (J1 - J11), 6 EB-type (J12 - J17), and 9 EA-type (J18 - J26) eclipsing binaries, whose types are qualified and classified by us using information such as the light curve patterns, periods, amplitudes, etc. After matching with the VSX and ASAS-SN catalogues, we find that 10 eclipsing binaries have been researched and classified by other studies \citep{18jaya, 18chen, 20chen}, and also find that there is no particular classification standard for the EA and EB type eclipsing binaries. Because the VSX catalogue includes variables discovered by a large number of surveys, some of EB-type binaries found in our work have been matched to the VSX catalogue, and we find that they were not only found in one survey and were determined to be of type EA or EB. Of course, these surveys also include the ASAS-SN survey. The eclipsing binaries that we have determined to be of type EB are consistent with the variable type given by ASAS-SN, so we have chosen the ASAS-SN determination criteria as a reference. Based on the work mentioned above, these remaining 16 are our newly discovered eclipsing binaries, which contain 7 EW, 2 EB, and 7 EA types. Among the 13 nights observations, we have detected some long-period eclipsing binaries, in which the longest -- J18 -- is up to about 7.2 days. However, due to night-time observations only, for some eclipsing binaries (J20, J22, J23, J26), our data cannot complete the phase diagram. Therefore, we have identified them as EA-type candidates except for those whose types have been determined by predecessors and can be identified by our visual method. The phase-folded light curve diagrams of all these eclipsing binaries are shown in Figure~\ref{fig:EB}. In addition, we have matched these eclipsing binaries with the open cluster \target members determined by \citet{18cant} using {\it Gaia} DR2, and obtained that 2 eclipsing binaries are the members of this cluster, one is an EW-type, and another is an EA-type eclipsing binary.

\begin{deluxetable*}{ccccccccccc}[!htp]
\tabletypesize{\small}
\tablewidth{0pc}
\tablenum{3}
\tablecaption{Specific information on eclipsing binaries in our observation.\label{tab:EB}}
\tablehead{
\colhead{ID} & \colhead{RA.} & \colhead{DEC.} & \colhead{$V$} & \colhead{P} & \colhead{A} & \colhead{$T_{eff}$} & \colhead{Type} & \colhead{Mem.} & \colhead{Reference} & \\
~ & \colhead{(degree)} & \colhead{(degree)} & \colhead{(mag)} & \colhead{(d)} & \colhead{(mag)} & \colhead{(K)} & ~ & ~ & ~ & }
\startdata
J1 & 108.7050954 & 13.7474155 & 14.63 & 0.704925149 & 0.31(1) & -- & EW & no & ASAS-SN\\  
J2 & 108.7378616 & 13.2368872 & 17.89 & 0.362369812 & 0.4(1) & -- & EW & no & our \\  
J3 & 109.8472352 & 13.6061456 & 18.24 & 0.31797351 & 0.7(2) & -- & EW & no & our\\  
J4 & 109.6147422 & 13.5698564 & 17.61 & 0.294312087 & 0.32(9) & -- & EW &no&our\\  
J5 & 109.502782 & 13.4805083 & 17.07 & 1.568406718 & 0.21(4) & -- & EW & no & our\\  
J6 & 109.4167725 & 13.7953294 & 16.77 & 0.696408296 & 0.11(3) & -- & EW & no & our\\  
J7 & 109.412816 & 14.3991508 & 17.7 & 0.306333395 & 0.42(6) & $5569\pm82$ & EW & no & C20\\  
J8 & 109.3495498 & 13.7812947 & 15.18 & 0.318030478 & 0.10(3) &--& EW &yes& our\\  
J9 & 109.2058456 & 13.7006874 & 14.97 & 0.664248285 & 0.54(1) & $6981\pm37$ & EW & no & C18\\  
J10 & 109.1749339 & 14.0861158 & 15.98 & 0.283403222 & 0.17(4) & -- & EW & no & C20\\  
J11 & 109.0838882 & 13.7072543 & 16.47 & 0.343961735 & 0.24(4) &--& EW &no&our\\  
J12 & 109.8432496 & 14.268068 & 14.79 & 0.290520284 & 0.54(1) & $4050\pm50$ & EB & no & C20\\  
J13 & 109.6709524 & 14.3419885 & 17.83 & 0.532057738 & 0.7(1) & -- & EB & no & our\\  
J14 & 109.4418925 & 14.2416662 & 15.81 & 0.575876016 & 0.54(3) & -- & EB & no & ASAS-SN\\  
J15 & 109.3781482 & 13.5799012 & 11.06 & 0.658875512 & 0.255(8) & $6406\pm21$ & EB & no & ASAS-SN\\  
J16 & 109.0725927 & 13.696204 & 15.31 & 0.502755629 & 0.61(2) & -- & EB & no & our\\  
J17 & 109.0658004 & 13.8280396 & 12.65 & 0.581627848 & 0.613(8) & -- & EB & no & ASAS-SN\\  
J18 & 109.6309312 & 13.9601402 & 15.55 & 7.206777787 & 1.40(6) & -- & EA & no & ASAS-SN\\  
J19 & 109.454682 & 14.3508626 & 17.80 & 0.549391707 & 1.0(3) & -- & EA & no & our\\  
J20 & 109.2499035 & 14.2216055 & 18.21 & 3.024331909 & 0.5(2) & -- & EA & no & our\\  
J21 & 109.1642519 & 13.8676077 & 17.19 & 0.913374903 & 0.80(8) & -- & EA & yes & our\\  
J22 & 109.1448447 & 13.7156306 & 14.69 & 1.545501416 & 0.11(1) & $6064\pm24$ & EA & no & our\\  
J23 & 109.0751204 & 14.0879028 & 14.66 & 2.562004676 & 0.18(2) & -- & EA & no & our\\  
J24 & 109.0422302 & 13.8363867 & 14.15 & 3.4220256 & 0.84(2) & -- & EA & no & ASAS-SN\\  
J25 & 108.9081585 & 13.6208906 & 17.04 & 0.797580838 & 0.59(5) & -- & EA & no & our\\  
J26 & 108.8794118 & 14.0339731 & 16.93 & 3.44607673 & 0.30(4) & -- & EA & no & our\\  
\enddata
\tablecomments{Column 1: variable stars' ID. Column 2 and 3: right ascension and declination (J2000). Column 4: the magnitude of variable stars in $V$ band. Column 5: the main period of variable stars. Column 6: the amplitude of folded light curves calculated by PERIOD04. Column 7: relatively accurate effective temperature from LAMOST DR7. Column 8: different types of variable stars. Column 9: membership of variable stars provided by \citet{18cant}. Column 10: reference of variables detection. ASAS-SN, C18 and C20 indicate that the variables were found by \citet{18jaya}, \citet{18chen} and \citet{20chen}, respectively.}
\end{deluxetable*}

\begin{figure}[!htp]
\begin{center}
\includegraphics[angle=0, width=1.0\columnwidth]{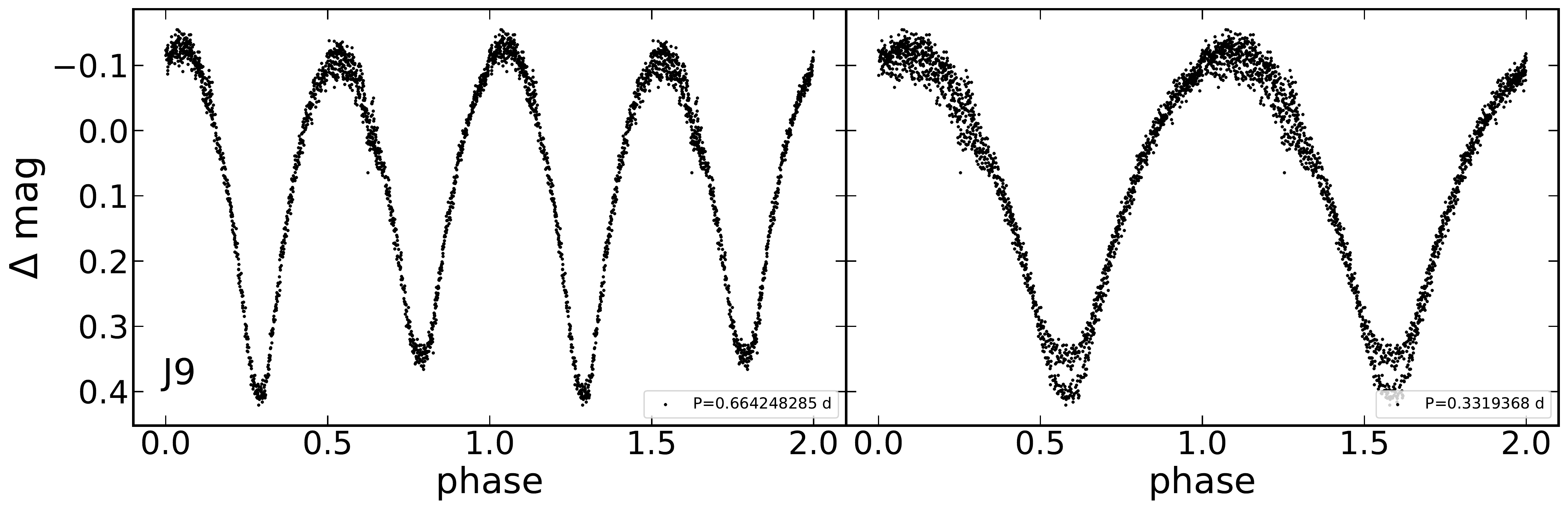}
\caption{The different phase-folded light curves for J9. left panel: the light curve of J9 with the period given by us. Right panel: the light curve with the period given by \citet{18chen} based on our observation data.}
\label{fig:J9}	
\end{center}
\end{figure}

On account of cross-match with other studies, \citet{18chen} identified variable J9 as an RR Lyrae variable in their study, they determined the period to 0.3319368 d, while we consider it as an eclipsing binary with the period of 0.664248285 d. From Figure~\ref{fig:J9} we can see that, the period given by \citet{18chen} to determine it as an RR Lyrae variable is approximately half of the period given by our determination of it as an EW-type eclipsing binary, and then we used the period given by \citet{18chen} to phase collapse our data to obtain the comparison plot in the right panel of Figure~\ref{fig:J9}. It's obvious to see that there is a gap between the primary and secondary maximum of the light curve. Moreover, the light curve of J9 does not conform to the sharp brightening and slow dimming nature of the RR Lyrae variable stars. Thus, we can determine that the variable J9 is an EW-type eclipsing binary with two components sharing a common Losch geometry envelope but with different temperatures based on the unequal height of its primary and secondary minimum moment. 

\subsection{Pulsating Stars} \label{sub:PS}
\begin{figure}[!htp]
\begin{center}
\includegraphics[angle=0, width=0.8\columnwidth]{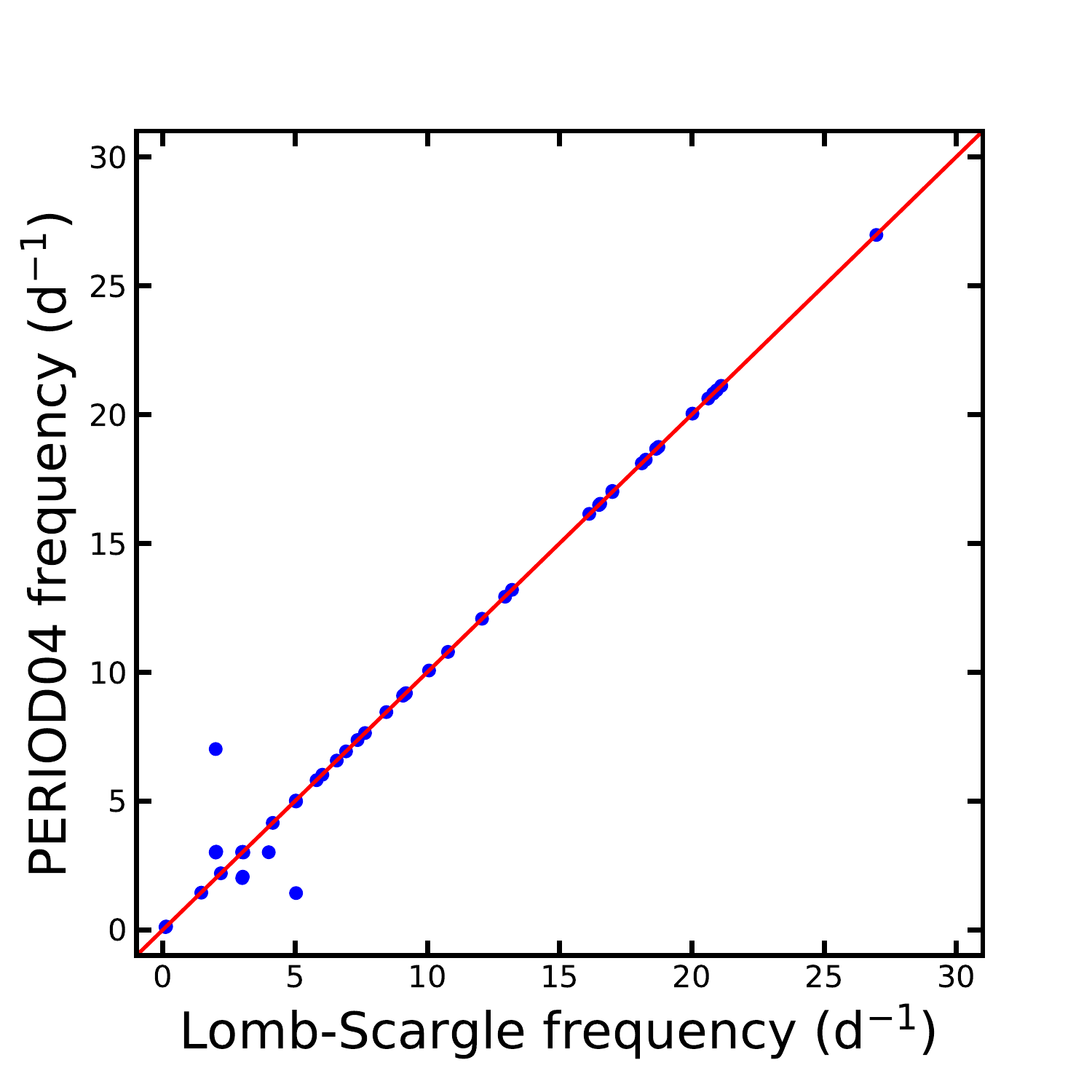}
\caption{Comparison of PERIOD04 and Lomb-Scargle extracted pulsating variable star frequencies.}
\label{fig:lsp}	
\end{center}
\end{figure}

\begin{figure*}[!htp]
\begin{center}	
\includegraphics[angle=0, width=1.8\columnwidth]{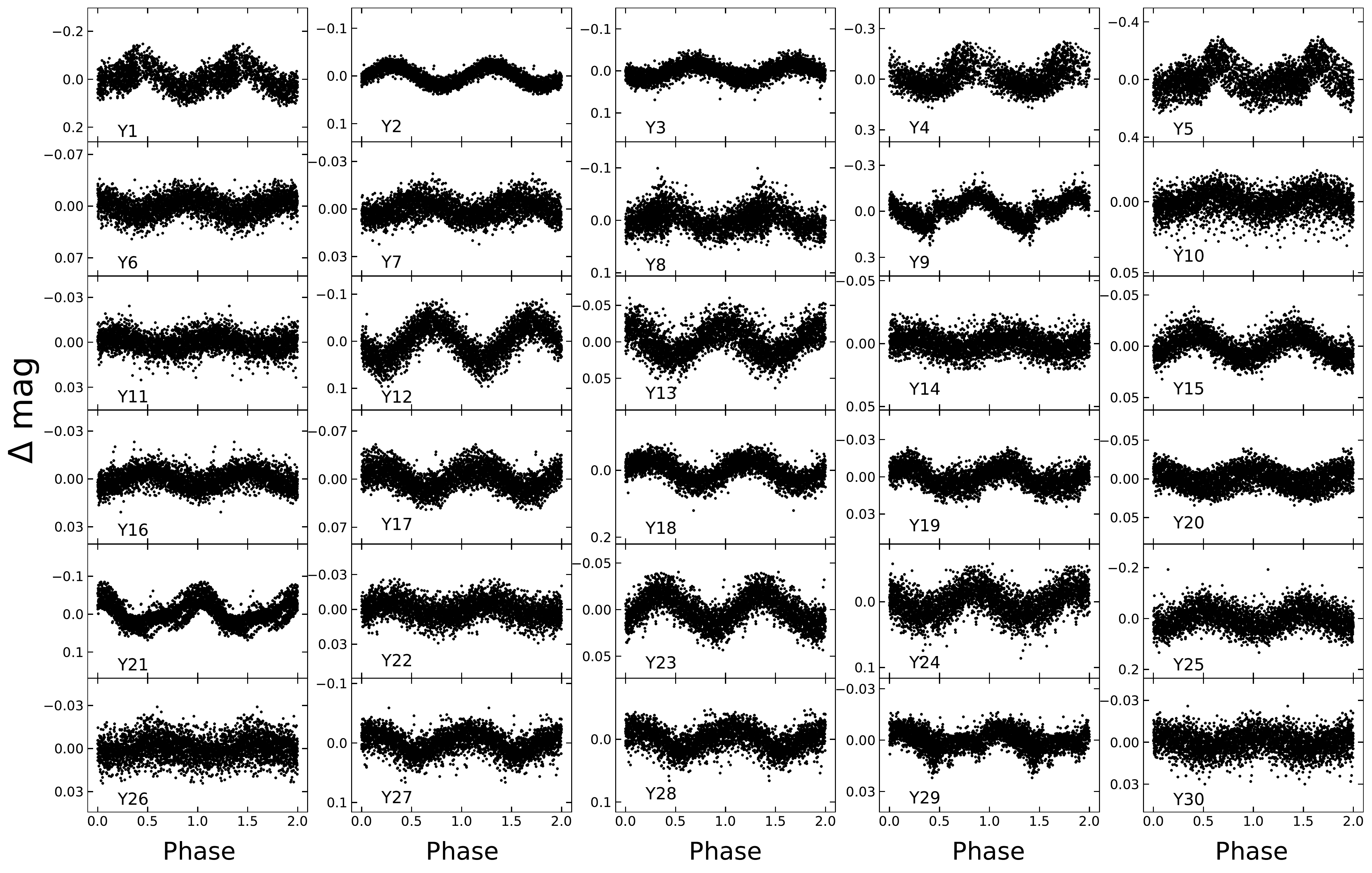}
\caption{Same as Figure~\ref{fig:EB}, but for pulsating stars of Y1-Y30.}
\label{fig:PS}
\end{center}
\end{figure*}

\begin{figure*}[!htp]
\begin{center}
\includegraphics[angle=0, width=1.8\columnwidth]{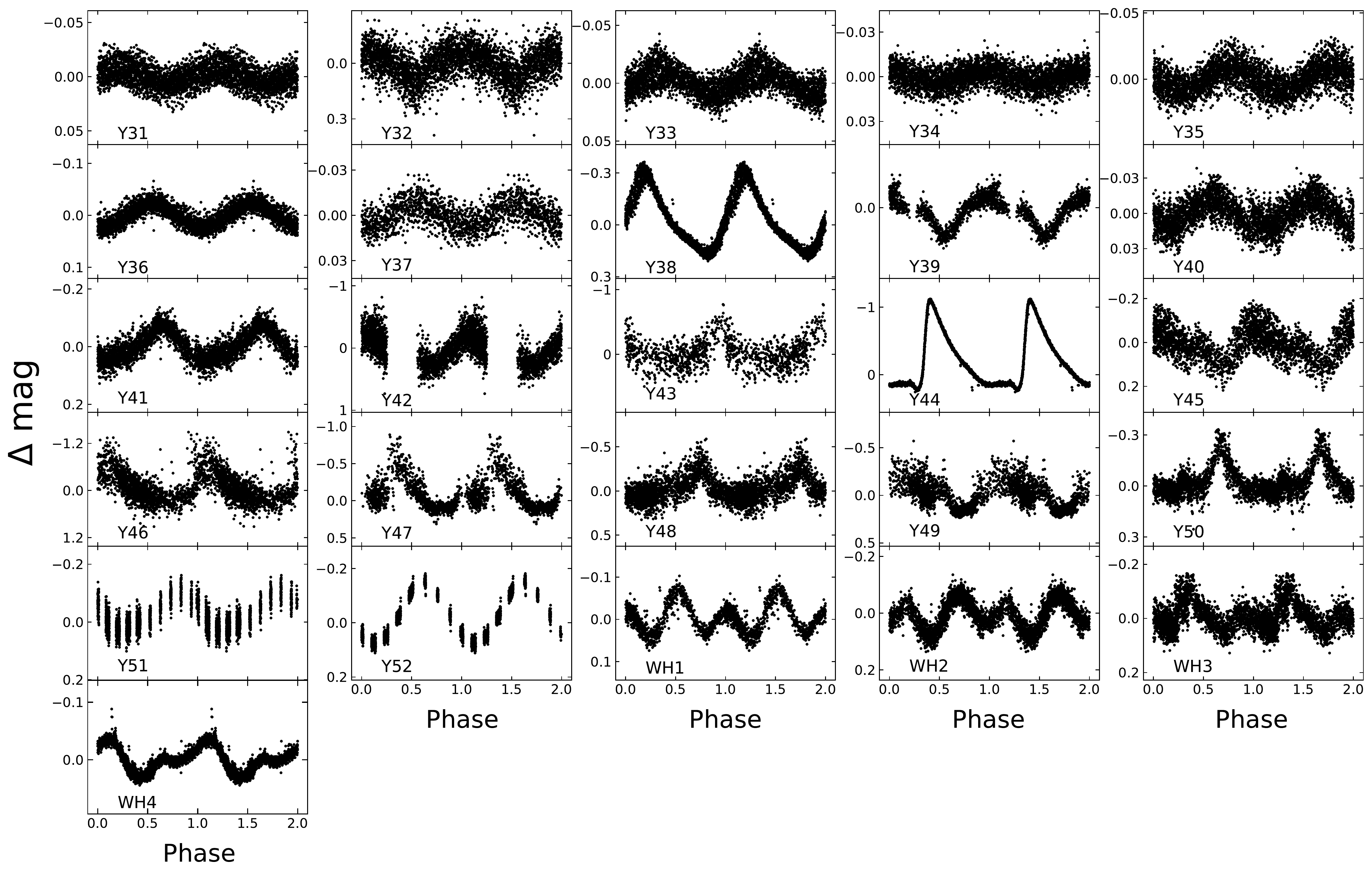}
\caption{Light curves of pulsating stars Y31-Y42 and other type variables W1-W4.}
\label{fig:PSOT}
\end{center}
\end{figure*}

Except for the eclipsing binaries mentioned above, we also find other $52$ pulsating variable stars with significant light curve variations. Referring to \citet{05duprmn} and \citet{05dupraa}, \citet{10aert} provided an approximate range of period, amplitude, and effective temperature for different types of pulsating stars, which we can use to give a detailed classification of pulsating variable stars in relation to their light curve properties \citep{11uytt}. To make a more accurate determination of the variable star types, we obtain some information especially effective temperature by matching the coordinates of these pulsating variable stars with LAMOST DR7. Since most pulsating variable stars have multi-mode pulsation modes, we calculated the fundamental and harmonic frequencies of each pulsating star using Period04 \citep{05lenz}. There is also a distinction between fundamental and harmonic frequencies of pulsating variable stars, and we generally take the frequency with the highest signal-to-noise ratio as the fundamental frequency to calculate the main period of the pulsating variable star as its intrinsic period. To verify the accuracy of all the pulsating stars' primary frequencies, then we performed fundamental frequency analyses for all time-series data in $V$ band by using the Lomb-Scargle periodogram \citep{15vand} of the Astropy package \citep{13astr,18astr}. The results are shown in Figure \ref{fig:lsp}, and both of them have relatively good agreement. All of their information is listed in Table~\ref{tab:PS}. Using these main periods extracted from fundamental frequencies, we show the phase folding diagrams of these $52$ pulsating variable stars in Figures~\ref{fig:PS} and \ref{fig:PSOT}.

Finally, we have classified these $52$ pulsating variable stars into different types: 38 $\delta$\,Scuti (Y1 - Y38), 3 $\gamma$\,Dor(Y39 - Y41), 9 RR Lyrae variable stars (Y42 - Y50), and 2 Cepheid I type variable stars (Y51 - Y52). As for these 9 RR variable stars, we also classify them specifically into RRab and RRc types pulsating variable stars according to their periods, amplitudes, the morphology of their light curves and so on. Furthermore, by comparing the bulk light curves, we find that some of them are worth further studying. For instance, a high amplitude $\delta$\,Scuti (HADS) among the 38 $\delta$\,Scuti variable stars and the 3 $\gamma$\,Dor we determined, which can be used to study the different modes of radial and non-radial modal pulsations in stellar evolution \citep{00breg}. Meanwhile, classical Cepheids are the most important and accurate distance indicators for us to establish astronomical distance scales \citep{18chen}. RR Lyrae constitutes another useful distance indicator for tracking ancient environments of the structure of Milky Way \citep{13drak, 16gran}, as well as the neighborhood of the solar system \citep{98layd}. These two types of variable stars have also been identified in our work. For the pulsating variable stars, we have done the same catalog matching for VSX and ASAS-SN as for the preceding eclipsing binaries, finding that only 3 of the pulsating stars (Y21, Y38, Y44) in our list were identified as variables in ASAS-SN and our results are consistent with theirs \citep{18jaya}. We have done a member star match for these pulsating variable stars as well, similar to the previous Subsection~\ref{sub:EB}, and eventually matched 9 members with the catalogue of \citet{18cant}. All of the 9 pulsating star members of the open cluster \target are $\delta$\,Scuti pulsating variables.
 
\begin{deluxetable*}{ccccccccccc}[!htp]
\tabletypesize{\footnotesize}
\tablewidth{0pc}
\tablenum{4}
\tablecaption{Detailed information on pulsating stars in our observation.\label{tab:PS}}
\tablehead{\colhead{ID} & \colhead{RA.} & \colhead{DEC.} & \colhead{$V_{mag}$} & \colhead{P} & \colhead{A} & \colhead{$T_{eff}$} & \colhead{Type} &\colhead{Mem.} &\colhead{Reference} & \\
~ & \colhead{(degree)} & \colhead{(degree)} & \colhead{(mag)} & \colhead{(d)} & \colhead{(mag)} & \colhead{(K)} & ~ & ~ & ~ & }
\startdata
Y1 & 109.9108557 & 13.7159228 & 15,89 & 0.331521548 & 0.054(2) &--& $\delta$\,Scuti & no & our  \\ 
Y2 & 109.8728109 & 14.0712854 & 14.48 & 0.058868469&0.0205(2)&$7418\pm44$&$\delta$\,Scuti & no & our  \\ 
Y3 & 109.8470245 & 14.3053479 & 15.42 & 0.144456746 & 0.0181(4) &--& $\delta$\,Scuti & no & our  \\ 
Y4 & 109.768027 & 14.0796784 & 16.68 & 0.142582429 & 0.063(3) &--& $\delta$\,Scuti & no & our  \\ 
Y5 & 109.7150925 & 13.5388977 & 17.90 & 0.166207444 & 0.073(4) &--& $\delta$\,Scuti & no & our  \\  
Y6 & 109.5935811 & 13.744437 & 14.72 & 0.082845863 & 0.0085(4) &--& $\delta$\,Scuti & no & our  \\  
Y7 & 109.5237211 & 14.0054341 & 12.97 & 0.060482395 & 0.0042(2) &--& $\delta$\,Scuti & no & our  \\  
Y8 & 109.5186289 & 13.8422864 & 15.66 & 0.331221714 & 0.0160(6) &--& $\delta$\,Scuti & no & our  \\  
Y9 & 109.5036782 & 13.7302877 & 16.64 & 0.33123391 & 0.080(1) &--& $\delta$\,Scuti & no & our  \\  
Y10 & 109.4982053 & 13.80885 & 14.33 & 0.048050167 & 0.0039(2) &--& $\delta$\,Scuti & candidate & our  \\  
Y11 & 109.5012411 & 14.0079472 & 13.62 & 0.049919365 & 0.0032(2) &--& $\delta$\,Scuti & candidate & our  \\  
Y12 & 109.4411585 & 13.7474767 & 16.36 & 0.241134804 & 0.0397(7) &--& $\delta$\,Scuti & no & our  \\  
Y13 & 109.3908596 & 14.0971415 & 14.60 & 0.152285134 & 0.0219(8) &--& $\delta$\,Scuti & no & our  \\  
Y14 & 109.3186704 & 13.8888941&14.30&0.054809912&0.0043(2)& $7292\pm33$& $\delta$\,Scuti & yes & our  \\  
Y15 & 109.3142335 & 13.7722935 & 14.22 & 0.053576176 & 0.0135(3) &--& $\delta$\,Scuti & yes & our  \\  
Y16 & 109.2418059 & 13.768747 & 12.88 & 0.075796607 & 0.0035(1) &--& $\delta$\,Scuti & yes & our  \\  
Y17 & 109.2872054 & 14.2818177 & 14.78 &0.109619824&0.0145(4)&$7149\pm30$&$\delta$\,Scuti & no & our  \\  
Y18 & 109.2700543 & 14.0465042 & 16.23 & 0.172303837 & 0.0337(7) &--& $\delta$\,Scuti & no & our  \\  
Y19 & 109.2186735 & 13.760371 & 13.3 & 0.130980433 & 0.0073(2) &--& $\delta$\,Scuti & yes & our  \\  
Y20 & 109.2324957 & 13.9456429&13.39&0.060662342&0.0082(3)& $7374\pm17$ & $\delta$\,Scuti & yes & our  \\ 
Y21 & 109.2154808 & 13.9060147 & 13.01 & 0.110080136 & 0.0266(5)&$7287\pm17$&$\delta$\,Scuti & no & ASAS-SN\\ 
 Y22 & 109.2077096 & 13.7325699 & 13.98 & 0.061953548 & 0.0093(2)&--& $\delta$\,Scuti & yes & our  \\ 
Y23 & 109.2169443 & 13.6855261 & 14.43 & 0.053359577 & 0.0170(3) &--& $\delta$\,Scuti & yes & our  \\ 
Y24 & 109.2148984 & 14.3025137 & 16.07 &0.047762516&0.0175(5)&$7315\pm219$& $\delta$\,Scuti & no & our  \\ 
Y25 & 109.1399024 & 14.0107299 & 16.52 & 0.037081897 & 0.029(1) &--& $\delta$\,Scuti & no & our  \\ 
Y26 & 109.1111801 & 13.8594634 & 14.05 & 0.058717766 & 0.0031(2) &--& $\delta$\,Scuti & yes & our  \\ 
Y27 & 109.0953706 & 13.4844115 & 15.78 & 0.109163347 & 0.0147(5)&$5941\pm119$&$\delta$\,Scuti & no & our  \\ 
Y28 & 109.0953684 & 13.4844122 & 15.78 & 0.109163347&0.0148(4)&$5941\pm119$&$\delta$\,Scuti & no & our  \\ 
Y29 & 109.0793748 & 13.4891903 & 13.41 & 0.330042421 & 0.0042(2) &--& $\delta$\,Scuti & no & our  \\ 
Y30 & 109.080247 & 13.9584684 & 14.09 & 0.047367988 & 0.0037(2) &--& $\delta$\,Scuti & candidate & our  \\ 
Y31 & 109.0186492 & 13.4600758 & 11.73 & 0.055242417 & 0.0066(3) &$7249\pm189$&$\delta$\,Scuti & no & our  \\ 
Y32 & 108.9851145 & 13.4064989 & 17.87 & 0.135827969 & 0.063(2) &--& $\delta$\,Scuti & no & our  \\ 
Y33 & 108.9608518 & 13.5199721 & 13.40 & 0.099382485 & 0.0106(2)&$7218\pm34$&$\delta$\,Scuti & no & our  \\ 
Y34 & 108.9209026 & 13.5424117 & 13.51 & 0.108904205 & 0.0038(2) &--& $\delta$\,Scuti & no & our  \\ 
Y35 & 108.8926011 & 13.7851013 & 13.23 & 0.09272541 &0.0093(2)&$7219\pm15$&$\delta$\,Scuti & candidate & our  \\ 
Y36 & 108.8190455 & 13.470833 & 14.95 & 0.118382592 & 0.0219(3) &--& $\delta$\,Scuti & no & our  \\ 
Y37 & 109.2025478 & 13.7603428 & 13.79 & 0.04849899 & 0.0073(3) & $7418\pm58$&$\delta$\,Scuti & yes & our  \\ 
Y38 & 109.5389492 & 14.0900076 & 13.93 & 0.077362389 & 0.2005(8) &--& HADT & no & ASAS-SN  \\ 
Y39 & 108.6997067 & 14.1040814 & 14.42 & 0.333430555 & 0.0280(4) & $7032\pm30$ & $\gamma$\,Dor & no & our  \\ 
Y40 & 109.6759171 & 13.2210089 & 14.85 & 0.703078873 & 0.0114(3) & $7071\pm30$ & $\gamma$\,Dor & no & our  \\ 
Y41 & 109.513806 & 13.474619 & 16.53 & 0.695089343 & 0.0570(9) &--& $\gamma$\,Dor & no & our  \\ 
Y42 & 109.703044 & 13.9135448 & 18.28 & 0.498651049 & 0.245(6) &--& RRab & no & our  \\ 
Y43 & 109.4809065 & 13.7855212 & 17.08 & 0.485694946 & 0.19(1) &--& RRab & no & our  \\ 
Y44 & 109.4685443 & 14.3504307 & 14.36 & 0.456430404 & 0.475(5) & $6385\pm31 $& RRab & no & ASAS-SN\\ 
Y45 & 108.7428509 & 14.0129341 & 17.00 & 0.200895398 & 0.070(2) &--& RRc & no & our  \\ 
Y46 & 109.8974297 & 13.933293 & 18.35 & 0.199458274 & 0.41(1) &--& RRc & no & our  \\ 
Y47 & 109.675003 & 14.3738616 & 16.83 & 0.333775517 & 0.256(7) &--& RRc & no & our  \\ 
Y48 & 109.5392842 & 14.3699187 & 17.88 & 0.332466859 & 0.142(4) &--& RRc & no & our  \\ 
Y49 & 109.3758993 & 13.4016689 & 16.94 & 0.331063381 & 0.130(4) &--& RRc & no & our  \\ 
Y50 & 109.3098908 & 13.8720478 & 16.19 & 0.332434033 & 0.087(2) &--& RRc & no & our  \\ 
Y51 & 109.5133375 & 14.0605146 & 16.26 & 9.658083467 & 0.0609(9) &--& Cepheid I & no & our  \\ 
Y52 & 109.4736844 & 13.943555 & 14.65 & 7.978416787 & 0.1127(4) &--& Cepheid I & no & our  \\
\enddata
\tablecomments{The same information as Table~\ref{tab:EB} but for pulsating stars.}
\end{deluxetable*}

\subsection{Rotating Variables} \label{sub:other}
In the classification of pulsating variable stars given by \citet{10aert}, there is another category of periodic variable stars that cannot be classified based on their amplitudes, periods, temperatures, and light curves. The detailed parameters of these 4 variables with other categories are listed in Table~\ref{tab:OTH} and their phase diagrams are shown in the later order of Figure~\ref{fig:PSOT}. 

Therefore, we first match these 4 variable stars presenting two peaks of different heights with the VSX and ASAS-SN star catalogues, finding that 2 stars were determined to be of the variable star by previous studies, one (WH2) being a BY Draconis-type variable star \citep{20chen}, while another (WH4) being a ROT-type variable star \citep{18jaya}. BY Draconis-type variable stars with periods from a pinch of 1 to 120 days and amplitudes from several hundredths to 0.5 mag in $V$ band, are emission-line dwarfs of the dKe-dMe spectral type exhibiting quasi-periodic light variations. The variation of light is caused by the axial rotation of the star with varying degrees of the unevenness of surface brightness, spots or chromospheric activity \citep{20chen}. The ROT-type variable is classified by \citet{18jaya} due to their light curves showing evidence of rotational modulation. It differs from the above 3 BY Draconis-type stars in the morphology of the light curves, which may be due to the periastron material accompanying the rotation of the star.

Since the remaining two variable stars (WH1, WH3) show similar morphology to WH2 in terms of light transitions, we tentatively conclude that both of them are also BY Draconis-type variable stars. Given that they are both caused by star rotation regardless of whether they are BY Draconis-type or ROT-type variable stars, we call this class of variable stars rotating variable stars. The light variation of the pulsating variable stars is mainly due to their endogenous causes, while the light change of the rotating variables is due to their own rotation accompanied by spots, flares, periastron matter, etc. The 4 rotational variable stars (WH1 - WH4) have light curves that are due to their own rotation, which can be caused either by the star brightness or spots on their surface or by the peri-stellar material around the stars rotating with them.

\begin{deluxetable*}{ccccccccccc}[!htp]
\tabletypesize{\small}
\tablewidth{0pc}
\tablenum{5}
\tablecaption{Detailed information on rotating variable stars in our observation. \label{tab:OTH}}
\tablehead{\colhead{ID} & \colhead{RA.} & \colhead{DEC.} & \colhead{$V_{mag}$} & \colhead{P} & \colhead{A} & \colhead{$T_{eff}$} & \colhead{Type} &\colhead{Mem.} &\colhead{Reference} & \\
~ & \colhead{(degree)} & \colhead{(degree)} & \colhead{(mag)} & \colhead{(d)} & \colhead{(mag)} & \colhead{(K)} & ~ & ~ & ~ & }
\startdata
WH1 & 108.7934201 & 13.2461453 & 15.48 & 0.256759358 & 0.0379(7) & $5000\pm47$ & BY & no & our  \\
WH2 & 109.2705053 & 13.4954579 & 16.48 & 0.524295961 & 0.0556(9) & -- & BY & no & C20  \\
WH3 & 109.1498495 & 14.0871941 & 16.19 & 0.333037361 & 0.037(1) & -- & BY & no & our  \\
WH4 & 109.0146283 & 14.1139968 & 13.19 & 0.722455061 & 0.0238(2) & -- & ROT & no & ASAS-SN\\  
\enddata
\tablecomments{The same information as Table~\ref{tab:EB} but for rotating variable stars.}
\end{deluxetable*}

\subsection{Unclear Variable Stars} \label{sub:unk}
\begin{figure}[!htp]
\begin{center}
\includegraphics[angle=0, width=0.8\columnwidth]{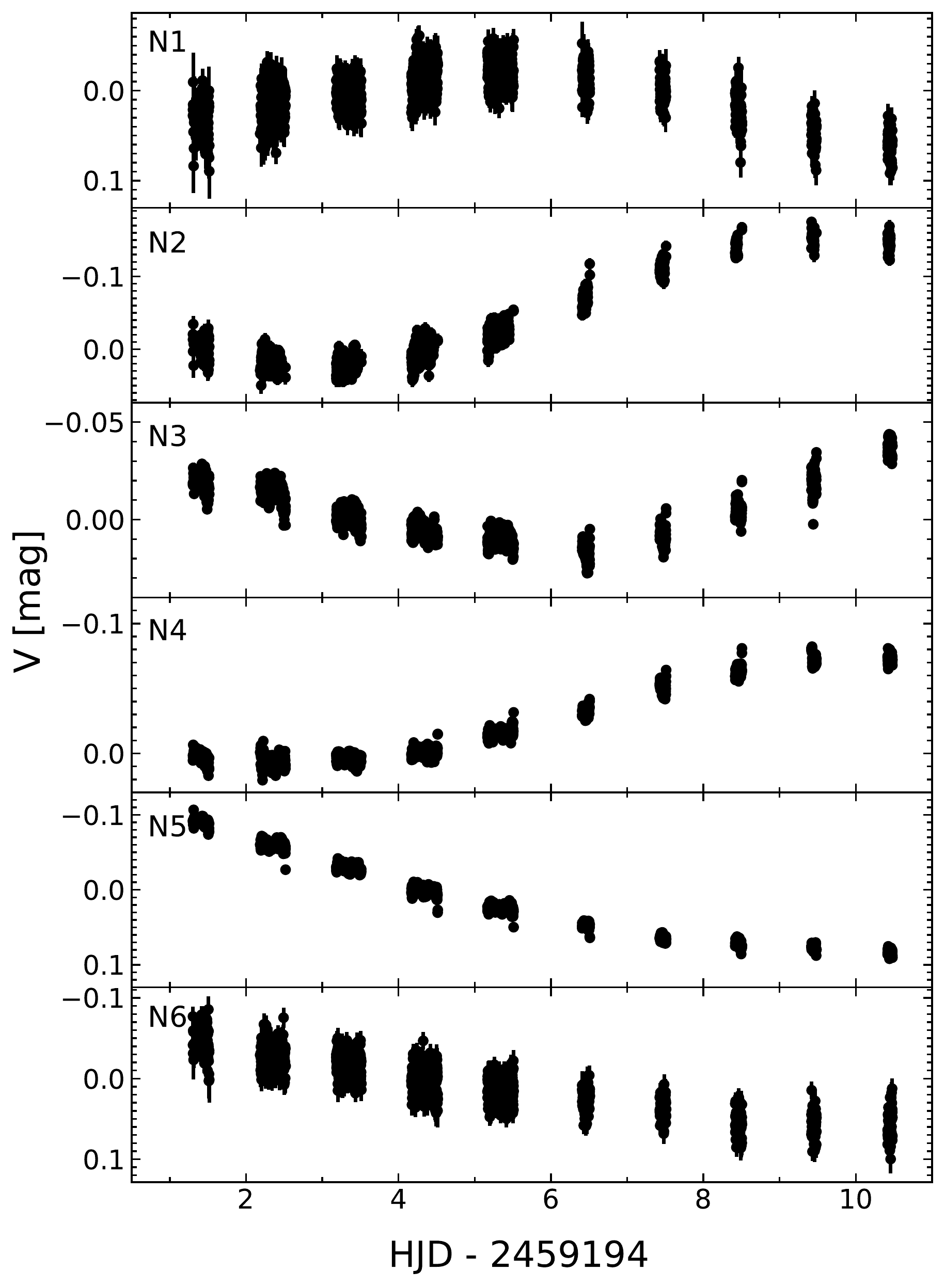}
\caption{The light curves of this 6 long period unclear variable stars which are listed in Table~\ref{tab:UN}.}
\label{fig:UN}	
\end{center}
\end{figure}

In addition to the periodic variables, we also detected 6 non-periodic variable stars (N1 - N6). It is difficult to classify these non-periodic stars only based on their light curves, so we list some of the parameters that can be obtained for these unclear variable stars as much as possible in Table~\ref{tab:UN}. The period of these unknown variable stars is relatively long, thus preventing us from obtaining the complete light curves of these long-period variable stars due to only a few days of observation data. From Figure~\ref{fig:UN} we can see that they show photometric variability, but owing to the variety of variable stars and their inability to achieve all conditions that exactly match the period, amplitude, and effective temperature information of a particular class of variable stars, and we tentatively classify them as unclear variable stars. 

By matching with the VSX and ASAS-SN catalogues, we obtained the type of one of the unclear variable stars, N5, which is an SR-type variable given by ASAS \citep{18jaya} with a period of up to about 465 days, a level we cannot reach with our short-term time-domain survey. The full name of the SR variable stars is Semi-Regular variable stars, which are cool giant evolved stars on the asymptotic giant branch. These are stars at the end of their lives, greatly expanded from birth, with very long pulsation periods, ranging from 10 to 1000 days. SR-type variables exhibit periodic variations, with interruptions in their pulsation periods leading to irregularities in their light curves. A large fraction of semi-periodic variable stars has two distinct modes of variability, which have long secondary periods superimposed on the shorter period variability \citep{18jaya}. At a later stage, we put these unclear variable stars together to study them as well, intending to find whether they belong to variable stars that are not identified as members of the open cluster NGC\,2355.

\begin{deluxetable*}{cccccccccc}[!htp]
\tabletypesize{\small}
\tablewidth{0pc}
\tablenum{6}
\tablecaption{Information on unclear variables in our observation.\label{tab:UN}}
\tablehead{\colhead{ID} &\colhead{RA.} &\colhead{DEC.} &\colhead{$V_{mag}$} & \colhead{P} &\colhead{$T_{eff}$} &\colhead{Type} &\colhead{Mem.} &\colhead{Reference} & \\
~ & \colhead{(degree)} & \colhead{(degree)} & \colhead{(mag)} & \colhead{(d)} & \colhead{(K)} & ~ & ~ & ~ & }
\startdata
N1 & 109.6418403 & 13.5294504 & 15.32 & -- & -- & unknown & no & our  \\ 
N2 & 109.5479767 & 14.3981865 & 13.65 & -- & -- & unknown & no & our  \\ 
N3 & 109.4011549 & 13.9683771 & 11.66 & -- & -- & unknown & no & our  \\ 
N4 & 109.3044602 & 14.1823171 & 11.46 & -- & -- & unknown & no & our  \\ 
N5 & 109.2825841 & 13.2909009 & 11.92 & 464.9215097 & -- & SR & no & ASAS-SN\\ 
N6 & 109.1929547 & 13.6178583 & 15.73 & -- & $4803\pm57$ & unknown & no & our  \\  
\enddata
\end{deluxetable*}

\section{Discussion}\label{sec:disc}
Overall, additional work has been done on the open cluster \target combined with the work of predecessors. Employing {\it Gaia} data, we have analyzed the nature of cluster variable member stars and obtained possible cluster member variable stars through a series of analyses. At the same time, based on the previous work done for this cluster, we have also refined the cluster parameters in conjunction with our work. 

\subsection{Membership of Variable Stars}\label{sub:mem}

\begin{figure}[!htp]
\begin{center}
\includegraphics[angle=0, width=0.9\columnwidth]{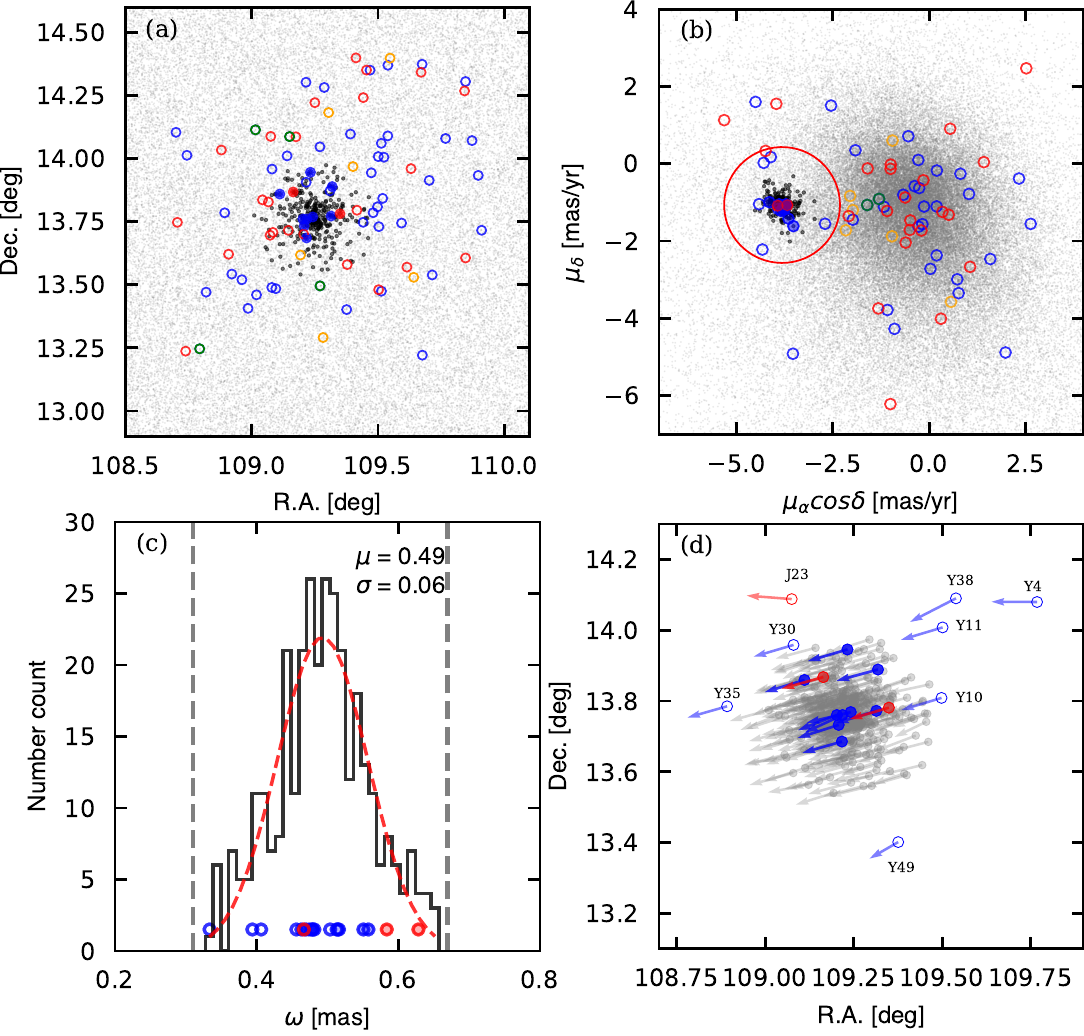}
\caption{(a): Spatial distribution for the cluster members and 88 variable stars, ($\alpha_{J2000}$ $\delta_{J2000}$). (b): proper motion distribution for them. (c): Histogram of parallax ($\omega$) for cluster members (black line) and variable stars located in the area of $3\sigma_{\omega}$ boundary (gray dashed lines). The histogram of members' parallax is fitted with a single Gaussian profile(red dashed line). (d): Spatial distribution for the cluster members and circled variable stars in (c) with their tangential velocities. Arrows point out the direction of tangential velocities for each star and arrow length is in proportion to the tangential speed. In top panels, light grey dots mean the complete samples of observation, and black dots represent 328 cluster members. Red, blue, green, and orange circles represent eclipsing binaries, pulsators, rotating variables, and unclear variables, respectively. Solid and hollow refer to whether they are matched members of NGC\,2355.}
\label{fig:mem}
\end{center}
\end{figure}

We cross-matched our variable stars with the member catalogue of \target reported by \citet{18cant} which contains $328$ members to check the membership of variable stars. $11$ variable stars were included in the cluster members, including $2$ eclipsing binaries and $9$ pulsating variables. In Figure \ref{fig:mem}, we show the homogeneousness of these variable cluster members in the spacial, proper motion, and parallax. 

Just as \citet{21tarr} pointed out that the members provided by \citet{18cant} were mainly focused on the inner parts of the clusters and unable to provide members in the peripheral regions of open clusters. Combined with proper motion and parallax data from {\it Gaia} DR2 \citep{18lind} and the members of \cite{18cant} in the core region, we reanalyze the membership for variable stars. Figure~\ref{fig:mem} (a) presents the spatial positions of the member stars of \target and the $88$ variable stars. The proper motion distributions of variable stars and members are shown in Figure \ref{fig:mem} (b). In proper motion space, we set a circular region centered on $(\mu_{\alpha}, \mu_{\delta}) = (-3.802, -1.086)$ mas yr$^{-1}$ with radius 1.5 mas yr$^{-1}$ as the selection criteria for variable stars. After excluding that located outside the criteria circle, $24$ variable stars remain, including $3$ eclipsing binaries and $21$ pulsating variables. 

To further purify the variable members, we apply the parallax criteria to select the variable member candidates from the $24$ remaining variables, which are shown in panel (c) of Figure \ref{fig:mem}. For the parallax distribution of the $328$ members of \citet{18cant}, a single Gaussian profile can be well fitted, we exclude variable stars with parallax $\omega$ greater than $3\sigma_{\omega}$ of the Gaussian distribution. After this selection, the variable member candidates retain $19$ stars, including $3$ eclipsing binaries and $16$ pulsating variables. 

In addition, we also employ proper motion parameters to perform the kinematic analysis for the selection boundaries, as shown in panel (d) of Figure \ref{fig:mem}. Panel (d) of Figure \ref{fig:mem} shows the spatial distributions and corresponding tangential velocities of the remaining $19$ variable stars and the cluster members. From the vector distribution of \target and variable stars, it can be seen that $4$ pulsating stars (V10, V11, V30, and V35) almost have the same direction and magnitude as that of the cluster members. 

We consider that V10, V11, V30, and V35 pulsating stars are very likely to be missing variable members of the cluster, which are located at the peripheral region of NGC\,2355. We present them as member star candidates in Table~\ref{tab:PS}.

\subsection{Stellar Parameters}\label{sub:para}
As we mentioned in the previous Section \ref{sec:O&D}, we not only performed $U, B, V, R, I$ multi-color metering on the standard star field, but also the open cluster NGC\,2355. We used the member catalogue of \citet{18cant} to match the cluster members with the stars in the field of view for which multi-color photometry was available and corrected and finally obtained 172 cluster members.

\begin{figure}[!htp]
\begin{center}
\includegraphics[angle=0, width=0.9\columnwidth]{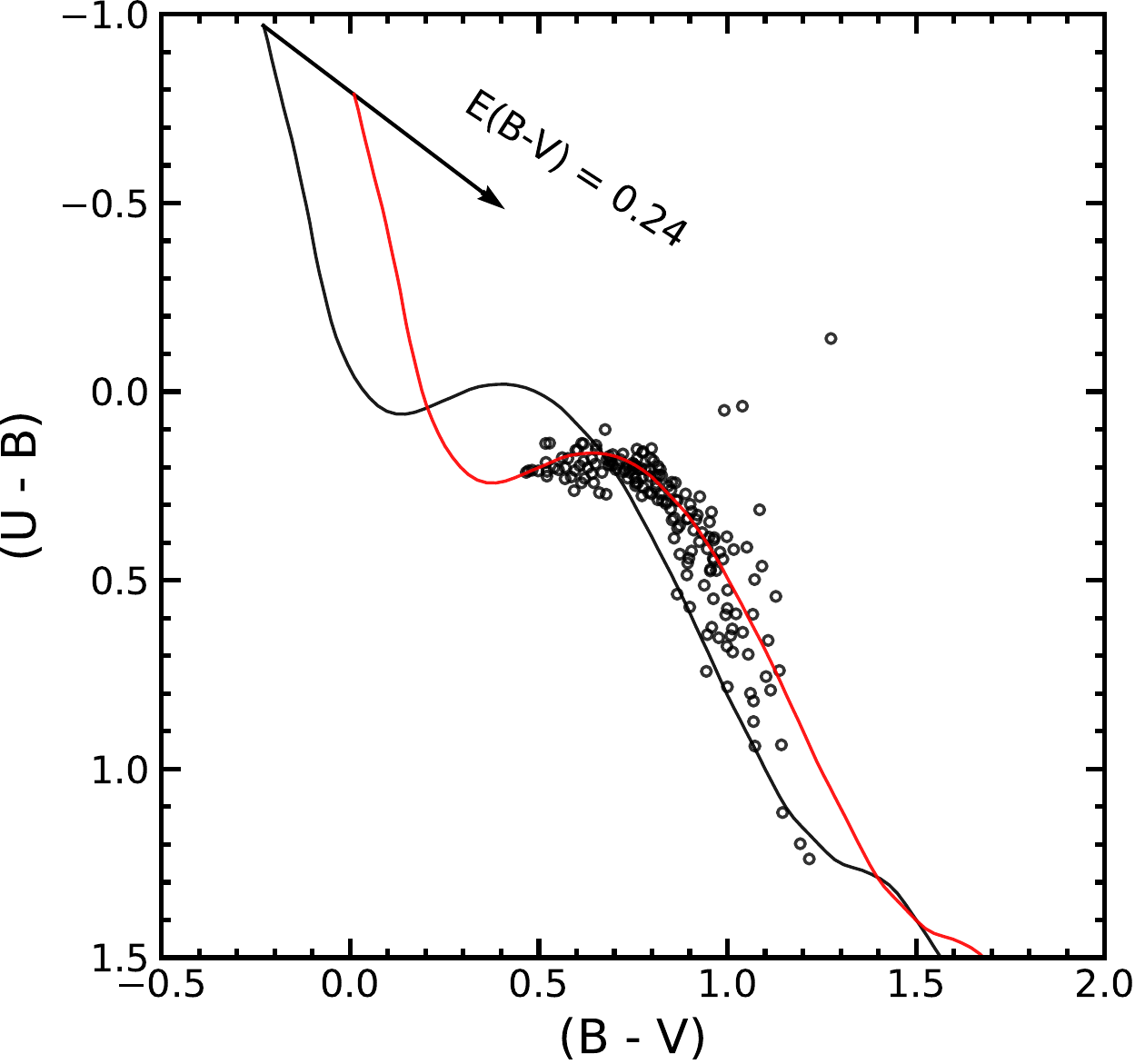}
\caption{The $(U-B)$ versus $(B-V)$ color-color diagram of the open cluster \target. The black solid curve line represents the locus of \citet{76turn, 79turn} ZAMS for solar metallicity. And, the red solid curve line is fitted to the black hollow circle by translating over the ZAMS with the condition given in the text.}
\label{fig:EBV}
\end{center}
\end{figure}

Reddening is one of the important basic parameters of open cluster, which can significantly affect other fundamental parameter determinations. The plots of classic two-color diagram (TCD, $U-B/B-V$) are very useful tools to estimate the reddening. We plotted the $(U-B)$ versus $(B-V)$ diagram in Figure \ref{fig:EBV} to estimate the reddening of NGC\,2355. The intrinsic zero-age main-sequence (ZAMS) from \citet{76turn, 79turn} is fitted by the reddening slope of $E(U-B)$/$E(B-V)$ as 0.76. Through fitting ZAMS to the cluster member, we obtained the $E(B-V) = 0.24\pm0.06$. The best fitting result is presented in Figure \ref{fig:EBV}.

\begin{figure}[!htp]
\begin{center}
\includegraphics[angle=0, width=0.8\columnwidth]{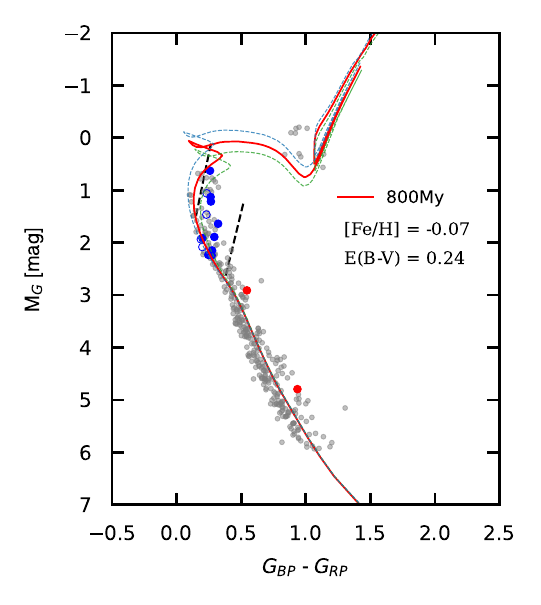}
\caption{CMD of open cluster \target with isochrone and the instability strip of $\delta$\,Scuti stars. Grey dots represent the 328 member stars of open cluster \target by \citet{18cant}. The PARSEC isochrones of 700, 800, and 900 Myr with metallicity [Fe/H]=-0.07 and extinction $A_{V} = 0.744$ are indicated by the blue dashed, green dashed, and red solid lines, respectively.} The approximate range of theoretical pulsation instability strip on the isochronous line is pointed by two black dashed lines. Different types of member variable stars are plotted with different symbols in the diagram the same as described in Figure \ref{fig:mem}.
\label{fig:CMD} 
\end{center}
\end{figure}

The CMD is usually used for the fundamental parameter estimations for open clusters, such as reddening, metallicity, distance, and age.  All the 328 members provided by \citep{18cant} are plotted in the CMD in Figure~\ref{fig:CMD}. To transform the observational magnitude $G$mag and colors ($G_{BP} - G_{RP}$) of each star to the absolute magnitude M$_{G}$ and intrinsic colors $(G_{BP} - G_{RP})$ \citep{18gaia}, distance and reddening are needed. We take the distance of $1902^{+447}_{-304}\,$pc estimated from \citet{18cant} for all the members, corresponding the distance modulus about $(m-M)= 11.4$ mag. For the extinction coefficients $A_{G}$, $A_{G_{BP}}$, and $A_{G_{RP}}$, we followed the transformation relation for Gaia bands: 
\begin{equation}
\begin{split}
        A_M/A_V= & c_{1M} + c_{2M}(G_{BP}-G_{RP}) + c_{3M}(G_{BP}-G_{RP})^2  \\
                 & +c_{4M}(G_{BP}-G_{RP})^3 + c_{5M}A_V + c_{6M}A_V^2  \\
                 & +c_{7M}(G_{BP}-G_{RP})A_V,
\end{split}
\label{eq:av}
\end{equation}
where M indicates the $G$, $G_BP$, or $G_RP$ band, and $c_{1...7M}$ belong to a set of coefficients defined in \citet{18gaiab} Table~1. 

In order to estimate the age of \target, we adopt the theoretical isochrones from PARSEC \citep{12bres} stellar evolution models to perform the CMD fitting. For the metallicity, we adopted the parameters provided by the high-resolution spectra of \citet{00soub}, \citet{11jaco} and \citet{15dona}, and averaged them to obtain a metal value [Fe/H] = $-0.07\pm0.01$ dex.
Using theoretical isochrones of different age (age $=700$, $800$, $900$ and $1000$ Myr) with [Fe/H] $= -0.07$ dex and $A_{V} = 0.744$ ($A_{V} = R_{V} \times E(B-V)$; $R_{V} = 3.1$). We adopted Equation (2) in \citet{19liu} to evaluate which of these isochrones shows the best fit. Since the $\overline{d}\,^{2}\,=\,0.0118$ for age $=\,800$\,Myr is less than the values $\overline{d}\,^{2}\,=\,0.0120$ and $0.0128$ for age $=\,700$ and $900$\,Myr, we consider $800\pm50$ Myr as the more reliable age for \target. Figure~\ref{fig:CMD} shows the isochrone fitting result.

Meanwhile, the instability region adopted from \citet{16jaff} and the $15$ variable members and member candidates are also presented in the CMD. It can be seen that all these pulsating star members are located in the instability region, which verifies the reliability of our variable star classification. 

\section{Summary}\label{sec:SM}
We have investigated and characterized an extensive variability search among stars in the field of the open cluster \target and its surrounding field by photometric observations with NOWT. We detected $88$ variable stars from $13$ nights photometric observations, containing $72$ new variable stars and $16$ known variables. They are classified into different types according to their main periods, amplitudes, the specific morphology of their light curves, effective temperature, etc. Finally, we discovered $26$ eclipsing binaries, including $11$ EW-type, $6$ EB-type, and $9$ EA-type eclipsing binaries. $52$ pulsating variable stars are also detected in this work, containing $38$ $\delta$\,Scuti stars which include 1 large-amplitude $\delta$\,Scuti (HADS), 3 $\gamma$\,Dor variables, 9 RR Lyraes, and 2 Cepheid I type variable stars. Besides, there are also 4 rotational type variable stars because their light variations are caused by rotation, and 6 long-period stars whose types are not clear. For those variable stars for which no type has been determined, we may need more observations later to identify and classify these unknown variable stars.

We also consult the cluster members established by \citet{18cant} and the database of {\it Gaia} DR2 to obtain the membership probability and the physical parameters of NGC\,2355. $11$ variable stars were determined as cluster members by \citet{18cant}, including $2$ eclipsing binaries and $9$ pulsating member stars. In addition to the $11$ variable members reported by \citet{18cant}, we find $4$ new cluster member candidates by detailed analysis. They are confirmed by the homogeneity of the position and kinematics of the star with the identified cluster members. We determined the reddening of \target to be $E($B$-$V$)= 0.24$ by TCD, and then combined the average of the metallicity given by previous authors using high-resolution spectra and the distance values given by \citet{18cant} to perform isochrones fitting for different age gradients. The best fit yields log(Age/yr) = 8.9, [Fe/H]=-0.07 dex, $m-M=11.4$ mag, and $E($B$-$V$)= 0.24$.

\normalem
\begin{acknowledgements}

We are grateful to an anonymous referee for valuable comments which help to improve the presentation. The authors acknowledge the National Natural Science Foundation of China under grant U2031204 and the science research grants from the China Manned Space Project with NO. CMS-CSST-2021-A08. The CCD photometric data of \target were obtained with the Nanshan One-meter Wide-field Telescope of Xinjiang Astronomical Observatory. This work has made use of data from the European Space Agency (ESA) mission {\it Gaia} (\url{https://www.cosmos.esa.int/web/gaia}), processed by the {\it Gaia} Data Processing and Analysis Consortium (DPAC, \url{https://www.cosmos.esa.int/web/gaia/dpac/consortium}). The Large Sky Area Multi-Object Fiber Spectroscopic Telescope (LAMOST) is a National Major Scientific Project built by the Chinese Academy of Sciences. Funding for the project has been provided by the National Development and Reform Commission. LAMOST is operated and managed by the National Astronomical Observatories, Chinese Academy of Sciences.

\end{acknowledgements}

\software{SExtractor \citep{96bert}, PERIOD04 \citep{05lenz}, IRAF \citep{86tody,93tody}, astropy \citep{13astr,18astr} }

\end{document}